\documentstyle[cite,axodraw,eqsecnum,tighten,preprint,aps,amstex]{revtex}

\def\chic#1{{\scriptscriptstyle #1}}
\newcommand{\db}{\rlap{$\partial$}/}

\newcommand{\slap}[1]{\rlap{$#1$}\hspace{0.3ex}/}

\newcommand{\A}{\slap{A}}
\newcommand{\cl}{{\cal L} }

\begin{document}

\preprint{\begin{tabular}{r} FTUV--01--0202\\ IFIC/01--4 \end{tabular}}

\draft

\title{Chiral fermions and gauge-fixing in five-dimensional theories}

\author{Joannis Papavassiliou and Arcadi Santamaria}

\address{\hfill{}\\
Departament de F\'{\i}sica Te\`orica and IFIC, Universitat de Val\`encia -- CSIC\\
 Dr. Moliner 50, E-46100 Burjassot (Val\`encia), Spain}
 
\maketitle

\begin{abstract}
We study in detail the issue of gauge-fixing in theories with one {\it
universal}  extra  dimension,  i.e.  theories  where  both  bosons  and
fermions display  Kaluza-Klein (KK) excitations.   The extra dimension
is compactified  using the  standard orbifold construction for a massless 
chiral fermion.   We carry
out the  gauge-fixing procedure at  the level of  the five-dimensional
theory  and  determine  the  tree-level  propagators  and  interaction
vertices needed  for performing perturbative calculations  with the
effective four-dimensional theory resulting     after    the
compactification.  The  gauge-independence of the  tree-level S-matrix
involving  massive KK modes  is  verified  using  specific
examples.  
In order to obtain massive fermionic zero modes one has to enlarge the theory 
by introducing a set of mirror fermions, a construction which is carried out in
detail. Finally, the
gauge-independence of the tree-level  S-matrix  involving the resulting  
new mass-eigenstates  is  proved   by  resorting  to  generalized  current
conservation equations.
\end{abstract}


\section{Introduction}

Field theories with  large extra dimensions have been  the focal point
of extensive  research in recent  years. 
Extra dimensions may  or may
not be accessible  to all known fields, depending  on the specifics of
the underlying, more fundamental theory.  In particular, a plethora of
variants have  been put forth,  where the usual Standard  Model fields
feel a smaller number of  extra dimensions compared to gravity, a fact
which  is  usually  enforced  by  means  of  standard  string-inspired
constructions
\cite{arkani-hamed:1998nn,arkani-hamed:1998rs,antoniadis:1990ew,
antoniadis:1994jp,bando:2000it,arkani-hamed:1998vp,dienes:1998sb,
dvali:1999cn,pilaftsis:1999jk,ioannisian:1999cw}.
The studies of  such theories presented so far  have mainly focused on
scenarios where  the fermions live  in the 4-d  (no KK modes)  and the
gauge-bosons in 5-d or higher, displaying KK 
modes \cite{pomarol:1998sd}.  The phenomenology
of such  models is  rich and  has been widely  explored in  the recent
literature   
(see for instance 
\cite{antoniadis:1994yi,antoniadis:1999bq,nath:1999mw,masip:1999mk,
delgado:1999sv,rizzo:1999br,carone:1999nz,nath:1999fs}).
On the other  hand, scenarios  where all  Standard Model
fields  live   in  higher  dimensions  are   less  explored
\cite{carone:1999nz}.  
Following
Appelquist, Chang, and Dobrescu
\cite{appelquist:2000nn},
we  will  refer  to the  former  type  of  extra
dimension(s) as  non-universal, and to  the latter type  as universal.
Models with  universal extra dimensions differ  from the non-universal
ones both theoretically  and phenomenologically.  From the theoretical
point  of  view,  accommodating   chiral  fermions  in  five 
dimensions  presents  well-known  subtleties,  and  necessitates
non-trivial  constructions. From the  phenomenological point  of view,
the most  characteristic feature of such theories  is the conservation
of the KK  number at each elementary interaction  vertex 
\cite{carone:1999nz,appelquist:2000nn}.
As a result,
and contrary to  what happens in the non-universal  case, the coupling
of any excited (massive) KK mode to two zero modes is prohibited.  One
immediate phenomenological  consequence is that  the lightest massive
gauge-bosons  corresponding  to  the  first excited  states  cannot  be
resonantly  produced  using normal  (zero-mode)  particles as  initial
sates; instead  they must be  pair-produced, a fact which  could place
them  beyond  the reach  of  the  next  generation of  colliders.   In
addition,  as has  been recently  shown, 
in the framework of theories with universal extra dimension(s)
one  is able to relax
relatively    tight   phenomenological    bounds   on the size of   the
compactification  scale obtained in  theories with
non-universal  extra  dimensions. For  example,  the  process $Z\to  b
\bar{b}$  is  reported \cite{appelquist:2000nn} 
to  furnish  less stringent  a  bound  in  the
universal  case than  in the  non-universal one,  first  considered in
\cite{papavassiliou:2000pq};  
in particular the  former yields  a 300  GeV bound,  to be
compared with  the 1TeV  bound obtained in  the latter.   Finally, the
conservation of the KK number  leads to the appearance of heavy stable
(charged   and  neutral)  particles,   which  may   pose  cosmological
complications (e.g. nucleosynthesis)\cite{appelquist:2000nn}.

Given the  interest in  such theories, one  would like to  establish a
well-defined calculational framework in order to accomplish a detailed
quantitative  study of  their  various phenomenological  implications.
One open issue in this  effort, which becomes relevant when attempting
to  extend  formal  considerations  to  actual  calculations,  is  the
question of how  to carry out correctly the  gauge-fixing procedure in
the context  of such theories.  
There are at  least three different cases where  the gauge fixing can
be of phenomenological importance.  Calculation of a simple $S$-matrix
element of massive fermions (non-conserved currents) involving a gauge
boson as an  intermediate particle are directly affected  by the gauge
fixing, already at tree level.   Moreover, one of the main features of
theories with extra dimensions is  the power-law 
running of the couplings,
which in turn offers the phenomenologically interesting possibility of
an  early unification \cite{dienes:1998vg,kakushadze:1998vr}.   
In  general, when  computing the  non-Abelian
contributions  to $\beta$  functions  non-trivial gauge  cancellations
take  place,  which  guarantee  that  the  resulting  expressions  are
gauge-independent.  Therefore, if one  wants to compute the entire set
of quantum  corrections due to  both fermionic and bosonic  loops, one
needs to  use the  correct propagators and  vertices, in order  not to
distort the aforementioned 
cancellations.  The gauge-fixing  procedure used here
as  well  as  most of  the  conclusions  regarding  the form  of  the
tree-level propagators carry over  to a non-Abelian context, augmented
by minor modifications when  elementary scalars with tree-level vacuum
expectation  values are  included.   In addition,  when computing  the
effective potential traditionally one  chooses a Landau type of gauge,
in  order to  eliminate a  certain  class of  diagrams
\cite{delgado:1998qr}. Therefore, the problem of defining the Landau gauge
in models with extra dimensions is a relevant one.

In  this paper we study  in detail the
following points in the context  of an Abelian gauge theory with one
extra universal dimension:

We  carry out the gauge-fixing  both {\it before}  and {\it after}
the compactification of the single extra universal dimension. We find that 
if the (covariant) gauge-fixing  is carried out before the compactification
the  Landau gauge cannot be  defined and that the unitary gauge cannot be
reached continuously from this type of gauges.
On the other  hand, if the gauge fixing is carried out after the 
compactification one arrives
at  a  Standard  Model-like  result  for the  various  propagators;  in
particular  both the  unitary  and the  Landau gauges  can be  reached
continuously. Notice in addition that 
(a) both constructions give rise to the same
propagators in the Feynman gauge, and (b) the unitary gauge
may be obtained directly before compactification
by means of a non-covariant, axial-gauge type of gauge
fixing~\cite{dienes:1998vg}.

Throughout  this paper  we have not  employed any  formal argument
which  guarantees that  one arrives  at the  same answer  for physical
observables  regardless of whether  one carries  out the  gauge fixing
before or  after the compactification,  especially in the  presence of
interactions.   Given the  importance of  this issue  we  have instead
resorted  to the  study  of explicit  examples  in the  context of  an
interacting theory, in order to establish whether at least some of the
typical features, known to be true for standard theories, persist.  In
order  to  accomplish  that,  we   have  carried  out  in  detail  the
construction necessary for defining fermions in 5 dimensions.  We have
adopted the orbifold compactification,  which is necessary in order to
eliminate from the spectrum the massless zero mode corresponding to the
fifth  component  of  the  gauge field,  which  is  phenomenologically
bothersome. If one adopts the orbifold
construction a  chiral structure  is introduced to a theory  which is
vector-like  at  the  level of the 5-d Lagrangian.   
Thus, after compactification  the fermionic zero modes are chiral and 
massless. On the other hand, the fermionic KK modes come in chiral pairs 
and can be combined to form Dirac fermions with masses which are 
integer multiples of the compactification scale. The coupling of the KK modes
to the bosons (vector and scalar) display the most general Lorentz structure 
in four dimensions. In this paper we do not consider problems related with 
chiral anomalies and 
assume that they cancel by adding the fermions with the right quantum numbers 
or the relevant Chern-Simons terms \cite{cheng:1999bg}. Notice that any
realistic model should reduce, at low energies, to the standard model in 
which chiral anomalies are canceled by a very particular choice of 
quantum numbers.

Using  the  boson propagators  derived  earlier we  
study tree-level $S$-matrix  elements involving  the massive KK  fermions as
external  particles.  Due  to  the KK  conservation  reflected in  the
elementary  vertices,  together  with  the  particular  mass  spectrum
mentioned above, we have shown that one arrives at the same answer for
the $S$-matrix elements considered,  
regardless of when the gauge-fixing
was carried  out.  

If one wants to obtain normal QED at low energies it is necessary to have
massive Dirac fermions as zero fermionic modes. However, the 
exercise of  endowing the zero modes with  mass is rather subtle.
If one was to assume that the Abelian theory we consider will
be eventually embedded  into a larger group, like the  $SU(2)\otimes U(1)$, 
one would expect to give masses to the zero modes by means of the standard
Yukawa coupling  of the fermions to  a scalar field,  which develops a
non-zero expectation  value through  the usual Higgs  mechanism.  This
standard  construction however  appears  to be  more  involved in  the
presence  of   an  orbifold-type of compactification,   
giving  rise  to a
subtle interplay of various field-theoretical mechanisms 
\cite{georgi:2000wb}.   
In addition, from  the theoretical  point of view,  one would like  to be
able to  deal with QED as  if it were a  self-contained theory without
having  to introduce  elementary scalars  in the  spectrum. Therefore,
in the simple Abelian case we will adopt the  construction whereby the task 
of giving masses
to the  fermionic zero modes is accomplished  through the introduction
of a set of mirror fermions, with opposite chirality properties. 
The proper definition
of mass-eigenstates after the zero modes have acquired masses requires
a  re-diagonalization  of  the   Lagrangian,  a  fact  which  in  turn
complicates the proof of the gauge-cancellations when computing simple
tree-level processes.  The  demonstration of these cancellations, once
accomplished,  furnishes  an  additional  non-trivial  check  for  the
robustness  of  both  the  gauge-fixing  procedure  and  the  orbifold
construction.   We have  indeed  shown that  these cancellations  take
place   by   virtue   of   general  expressions   reflecting   current
(non)conservation.

The paper is organized as follows: In section~\ref{sec:gauge-fixing}  we 
present the gauge fixing, which is carried out both before
and after the orbifold compactification.
In section~\ref{sec:chiral-fermions} we add interactions, complete 
the orbifold construction for fermions, and derive the elementary 
interaction vertices. 
In section~\ref{sec:gauge-invariance} we demonstrate 
with some simple examples the gauge-invariance of the
S-matrix, and that the two gauge-fixing procedures are equivalent,
at least at this basic level. In section~\ref{sec:massive-modes} we 
enlarge the spectrum in order to give masses to the fermionic zero modes
in a way compatible with the orbifold symmetry. We also check that the 
gauge symmetry of the S-matrix is preserved in the extended theory.
Finally, in section~\ref{sec:conclusions}
we present our conclusions. 

\noindent

\section{Gauge-fixing and tree-level propagators}
\label{sec:gauge-fixing}

In  this section  we will  carry out  the gauge-fixing  in  detail and
derive  the expressions for  the various  propagators.  First  we will
start from a five-dimensional  Lagrangian which is already gauge-fixed
before compactification,  and we will derive  the propagators obtained
after  the compactification  of the  fifth dimension  on  an orbifold,
following two different but equivalent procedures. Then we will derive
the propagators  starting from a five-dimensional  Lagrangian which is
not gauged fixed  before the compactification, and will  carry out the
gauge  fixing on  the four-dimensional  Lagrangian obtained  after the
compactification.
 
\subsection{The gauge-fixing before the compactification}

We will use a notation in which $M,N,\cdots =0,1,2,3,4$ denote 
five-dimensional indices and $\mu,\nu,\cdots = 0,1,2,3$ are four-dimensional
indices. The metric $g^{MN}$ is the standard five-dimensional Lorentz metric
with spatial signature for the extra dimension. We will denote the extra
spatial coordinate as $y$ and will write $x^M = (x^\mu,y)$ (so $x^4=-x_4=y$).
On the other hand, for the gauge fields we will write
$A_M(x,y) = (A_\mu(x,y),A_{\chic 5}(x,y))$, where $x$ denotes collectively the
four-dimensional coordinates.  

We start out with the free part ${\cal L}_{0}$
of the $QED_5$ Lagrangian 
and add a covariant gauge fixing term ${\cal L}_{gf}$. With the notation
introduced above we write the five-dimensional action as
\begin{equation}
S^{(5)} = \int d^4xdy ({\cal L}_{0}+{\cal L}_{gf})~,
\end{equation}
with
\begin{eqnarray}
{\cal L}_{0}+{\cal L}_{gf} &=& 
-\frac{1}{4}(\partial_{\chic M} A^{\chic N} 
-\partial_{\chic N} A^{\chic M})^2 
-\frac{1}{2a} (\partial_{\chic M} A^{\chic M})^2 \nonumber\\
 &=& \frac{1}{2} A^{\chic M}
\Bigg[ \Box_{\chic 5} g_{\chic M \chic N}-
\bigg(1-\frac{1}{a}\bigg) 
\partial_{\chic M} \partial_{\chic N} \Bigg]
A^{\chic N}\, ,
\label{gf1}
\end{eqnarray}
where in the second line we have carried out partial integrations, and
$\Box_{\chic 5} = \Box_{\chic 4}-\partial^2_y$
with $\Box_{\chic 4}=\partial_\mu\partial^\mu$.
Next assume that $y$ is compactified 
on a circle of radius $R$ with the points $y$ and $-y$ identified,
i.e. on an orbifold  
$S^1/{\mathbb Z}_2$. In general, fields even under the ${\mathbb Z}_2$ symmetry
will have zero modes which will be present in the low-energy
theory, whereas fields odd under ${\mathbb Z}_2$ will only have KK modes,
and their zero modes will disappear. 
For the case at hand,
we have that
$A_{\mu}$ transforms like $\partial_{\mu}$ (even under ${\mathbb Z}_2$),
whereas $A_{\chic 5}$ transforms like $\partial_y$ (odd under ${\mathbb Z}_2$).
Therefore, the  Fourier expansion of the fields has the form
\begin{eqnarray}
A_{\mu}(x,y) &=& \frac{1}{\sqrt{L}}
\Bigg[A_{\mu}^{(0)}(x)
+ \sqrt{2} \sum_{n=1}^{\infty} 
A_{\mu}^{(n)}(x)\cos (m_n y)\Bigg]
\nonumber\\
A_{\chic 5}(x,y) &=& \sqrt{\frac{2}{L}}\sum_{n=1}^{\infty} 
A_{\chic 5}^{(n)}(x)\sin (m_n y)~,
\end{eqnarray}
with
\begin{equation}
m_n = \frac{n}{R}\qquad {\mathrm and}\qquad L=\pi R\, .
\label{Mass}
\end{equation}
The above normalization of the  Fourier expansion
results in the canonical form for the kinetic terms. 
Following the standard KK construction, we arrive at
\begin{equation}
S^{(4)} = \int d^4x 
\sum_{n=0}^{\infty}
\bigg({\cal L}_{V}^{(n)}+{\cal L}_{S}^{(n)}+{\cal L}_{M}^{(n)}\bigg)~,
\end{equation}
with 
\begin{eqnarray}
{\cal L}_{\chic V}^{(n)} &=& 
\frac{1}{2} A_{\mu}^{(n)}
\Bigg[ (\Box_{\chic 4}+ m_n^2) g^{\mu\nu}-
\bigg(1-\frac{1}{a}\bigg) 
\partial^{\mu} \partial^{\nu}\Bigg]A_{\nu}^{(n)}\nonumber\\
{\cal L}_{\chic S}^{(n)} &=& - \frac{1}{2} A_{\chic 5}^{(n)}
\Bigg[\Box_{\chic 4} + \frac{1}{a}m_n^2 \Bigg] A_{\chic 5}^{(n)}\nonumber\\
{\cal L}_{\chic M}^{(n)} &=& -\frac{1}{2} \bigg(1-\frac{1}{a}\bigg)m_n
\Bigg[ A_{\mu}^{(n)}\partial^{\mu} A_{\chic 5}^{(n)}
-  A_{\chic 5}^{(n)}\partial^{\mu} A_{\mu}^{(n)}\Bigg]~, 
\label{LGF}
\end{eqnarray}
where the subscripts stand for
vector (V), scalar (S), and mixed (M).
Clearly, ${\cal L}_{\chic M}^{(0)}=0$, and  
${\cal L}_{\chic S}^{(0)}=0$ because  $A_{\chic 5}^{(0)}=0$.  
Notice moreover that the mass term for 
the  $A_{\chic 5}^{(n)}$ is multiplied by
the inverse of the gauge-fixing parameter,  
and that the mixing term vanishes in the
Feynman gauge, $a=1$ .

In order to determine the propagators one must re-diagonalize the
Lagrangian. 
The quadratic part given above may be cast in the form
\begin{equation}
\Bigg(A_{\mu}^{(n)} A_{\chic 5}^{(n)}\Bigg) 
\left(
\begin{array}{cc}
(\Box_{\chic 4} + m_n^2)g^{\mu\nu} - (1-1/a)\partial^{\mu}\partial^{\nu} 
& -(1-1/a)m_n \partial^{\mu}\\
(1-1/a)m_n \partial^{\nu} &  -\Box_{\chic 4} - m_n^2/a
\end{array}\right) 
\left(\begin{array}{c} A_{\nu}^{(n)}\\  
A_{\chic 5}^{(n)}
\end{array}\right)~.
\end{equation}
In the momentum space the $2\times 2$ matrix given above, to be denoted by $D_n$,
assumes the form 
\begin{equation}
D_n = \left(
\begin{array}{cc}
(m_n^2-q^2)g^{\mu\nu} + (1-1/a)q^{\mu}q^{\nu} 
& i(1-1/a)m_n q^{\mu}\\
-i(1-1/a)m_n q^{\nu} &  q^2 - m_n^2/a
\end{array}\right)~.
\end{equation}
Next we must find the inverse of the matrix; using the condition  
$D_n D^{-1}_n = 1$, and using the parametrization 
\begin{equation}
D^{-1}_n = \left(
\begin{array}{cc}
 \Delta_{\mu\nu}^{(n)}(q) 
& \Delta_{\mu}^{(n)}(q)\\
-\Delta_{\nu}^{(n)}(q) & \Delta^{(n)}(q)
\end{array}\right)
\end{equation}
we find
\begin{eqnarray}
i\Delta_{\mu\nu}^{(n)}(q) &=&  \frac{-i}{q^2-m_n^2}
\Bigg[g_{\mu\nu}  
- \frac{(1-a)q_{\mu}q_{\nu} }{q^2-m_n^2}\Bigg] \nonumber\\
i\Delta_{\mu}^{(n)}(q) &=&  \frac{m_n(1-a)}{(q^2-m_n^2)^2}q_{\mu} 
\nonumber\\
i\Delta^{(n)}(q)&=& i\Bigg[\frac{1}{q^2-m_n^2} + 
\frac{m_n^2(1-a)}{(q^2-m_n^2)^2}\Bigg]~, 
\label{Props}
\end{eqnarray}
shown in Fig.~1a -- Fig.~1c .
Notice that there is no value for $a$ such that one would recover a 
unitary gauge type of propagator. In particular, the
choice ($a=0$) does not decouple the scalar sector and
does not lead to the unitary gauge type of spectrum. 
In addition, as is evident from the first expression
in Eq.~(\ref{Props})
the Landau gauge, i.e. a gauge 
where $\Delta_{\mu\nu}^{(n)}(q)$ is proportional
to the usual transverse structure
$(g_{\mu\nu} -q_{\mu}q_{\nu}/q^2)$ 
 cannot be defined.
Instead, as we will see in the last sub-section, 
both the unitary gauge as well as the Landau gauge
may be reached
if the gauge fixing is performed  
at the level of the four-dimensional 
Lagrangian obtained after carrying out the compactification
of the extra dimension.

\subsection{The Dyson summation}

Next we will show that one arrives at exactly the expressions
for the propagators given in Eq.~(\ref{Props}) 
if one
treats the mixing term as an interaction,  
and carries out the
Dyson summation.

To that end let us invert naively the
quadratic parts of $A_{\chic 5}^{(n)}$ and $A_{\chic 5}^{(n)}$ and obtain the
tree-level propagators $d_{\mu\nu}^{(n)}(q)$ and $d^{(n)}(q)$,
respectively, given by \cite{delgado:1998qr} (Fig.~2a and Fig.~2b)
\begin{eqnarray}
id_{\mu\nu}^{(n)}(q) &=& \frac{-i}{q^2-m_n^2} 
\Bigg[{g_{\mu\nu}} 
- \frac{(1-a)q_{\mu}q_{\nu} }{q^2-a m_n^2}\Bigg]
\nonumber\\
id^{(n)}(q)  &=& \frac{i}{q^2-m_n^2/a}
\label{dProps}~.
\end{eqnarray} 
The mixing interaction is
given by the vertex (the momentum
flow convention is shown in Fig.~2c) 
\begin{equation}
{\cal V}_{n}^{\mu} = \bigg(1-\frac{1}{a}\bigg)m_n q^{\mu}\, .  
\end{equation}
The basic quantity appearing in the Dyson sum is (no sum over $n$)
\begin{eqnarray}
K_n &=&  d^{(n)} {\cal V}^{\mu}_{n} 
{\cal V}^{\nu}_{n} d_{\mu\nu}^{(n)}
\nonumber\\
&=&  
-a \bigg(1-\frac{1}{a}\bigg)^2 q^2 m_n^2
 \Bigg[\bigg(q^2-\frac{m_n^2}{a}\bigg) 
\bigg(q^2-a m_n^2\bigg)\Bigg]^{-1}~.
\end{eqnarray}
Then $\Delta_{\mu\nu}^{(n)}(q)$ is given by (Fig.~2d)
\begin{eqnarray}
i\Delta_{\mu\nu}^{(n)} &=& id_{\mu\nu}^{(n)} +
id^{(n)} \bigg({\cal V}^{\rho}_n d_{\mu\rho}^{(n)}\bigg)
\bigg({\cal V}^{\sigma}_n d_{\nu\sigma}^{(n)}\bigg)
\Bigg[1+ \sum_{\ell=1}^{\infty} (K_n)^{\ell}  \Bigg]\nonumber\\
&=& id_{\mu\nu}^{(n)} + 
\Bigg(\frac{i(1-a)^2 m_n^2}{(q^2-m_n^2/a)(q^2-a m_n^2)^2} 
\Bigg)\Bigg[ \frac{1}{1-K_n}\Bigg] q_{\mu}q_{\nu}
\nonumber\\
&=& id_{\mu\nu}^{(n)}+
\frac{i(1-a)^2 m_n^2}{(q^2-m_n^2)^2 (q^2-a m_n^2)} 
q_{\mu}q_{\nu}
\label{SDPropa}~.
\end{eqnarray}
Similarly, for $\Delta^{(n)}(q)$ (Fig.~2e) we have
\begin{equation}
i\Delta^{(n)} = id^{(n)}+id^{(n)}K_n
\Bigg[1+ \sum_{\ell=1}^{\infty} (K_n)^{\ell}\Bigg]=
\frac{id^{(n)}}{1-K_n}=
id^{(n)}\frac{(q^2-a m_n^2)(q^2- m_n^2/a)}{(q^2-m_n^2)^2}
\label{SDPropb}~.
\end{equation}
It is elementary to verify, by using the expressions for  
$d_{\mu\nu}^{(n)}$ and $d^{(n)}$  given in Eq.~(\ref{dProps}),
that the right-hand sides of Eq.~(\ref{SDPropa}) 
and Eq.~(\ref{SDPropb}) indeed reduce to those reported 
to the expressions for $\Delta_{\mu\nu}^{(n)}$ and 
$\Delta^{(n)}$, respectively, given 
in Eq.~(\ref{Props}). 
The mixing term $\Delta_{\mu}^{(n)}$
can be obtained in a similar fashion. 

\subsection{The Gauge Fixing after the compactification}

In this subsection we will carry out the 
KK construction first, and then we will do the gauge-fixing
directly in the 4-d theory. 
The result of 
compactifying the $y$  
before adding a gauge fixing term may be 
worked out straightforwardly, or equivalently
gleaned off directly from Eq.~(\ref{LGF})
by setting $a \to \infty$ . We repeat that this limit does
not amount to the unitary gauge; in particular, due to the
mixing term which does not vanish, we do not arrive at
the Proca Lagrangian for the $n$-th massive gauge boson. 
Instead we have
\begin{eqnarray}
\widehat{{\cal L}}_{\chic V}^{(n)} &=& 
\frac{1}{2} A_{\mu}^{(n)}
\Bigg[ (\Box_{\chic 4} + m_n^2) g^{\mu\nu}- 
\partial^{\mu} \partial^{\nu}\Bigg]A_{\nu}^{(n)}\nonumber\\
\widehat{{\cal L}}_{\chic S}^{(n)} &=& -\frac{1}{2} A_{\chic
5}^{(n)}\Box_{\chic 4} A_{\chic 5}^{(n)}
\nonumber\\
\widehat{{\cal L}}_{\chic M}^{(n)} &=& m_n
A_{\chic 5}^{(n)}\partial^{\mu} A_{\mu}^{(n)} ~.
\end{eqnarray}
Then we add a gauge-fixing term which corresponds to the
generalization of the usual $R_\xi$ gauge-fixing, used 
in the electroweak sector of the Standard Model, 
\begin{equation}
\widehat{{\cal L}}^{(n)}_R = 
-\frac{1}{2\widehat{a}_n}\bigg(\partial^{\mu} A^{(n)}_{\mu} +
\widehat{a}_n m_n A^{(n)}_{\chic 5}\bigg)^2~.
\end{equation}
Notice that we introduce an infinity of arbitrary
gauge fixing parameters $\widehat{a}_n$; thus one can carry
out the gauge-fixing independently for every single gauge
boson, exactly as in the Standard Model
case, when one may choose three
completely independent 
gauge-fixing parameters $\xi_{\gamma}$,
$\xi_{\chic Z}$, and $\xi_{\chic W}$ for the photon, the $Z$-boson,
and the $W$-boson, respectively. 
By construction this gauge fixing removes the tree-level 
mixing and gives rise to the following two propagators
\begin{eqnarray}
\widehat{\Delta}_{\mu\nu}^{(n)}(q) &=&  
\frac{-i}{q^2-m_n^2} \Bigg[g_{\mu\nu}  
- \frac{(1-\widehat{a}_n)q_{\mu}q_{\nu} }
{(q^2-\widehat{a}_n m_n^2)}\Bigg] 
\nonumber\\
\widehat{\Delta}^{(n)}(q)&=& \frac{i}{q^2-\widehat{a}_n m_n^2}~.  
\end{eqnarray}
Clearly, in the Feynman gauge ($\widehat{a}_n =1$) 
we recover the same propagators as
in the Feynman gauge of the previous subsection, 
$a=1$. But in addition, now
the Landau gauge ($\widehat{a}_n =0$)
gives indeed a transverse propagator
for the massive gauge boson, and a massless scalar propagator,
the analogue of a massless would-be Goldstone boson. In addition,
in the limit  $\widehat{a}_n \to \infty$ one recovers the standard 
unitary propagator, i.e.
\begin{equation}
U_{\mu\nu}^{(n)}(q)= \frac{-i}{q^2-m_n^2}\Bigg[g_{\mu\nu} - 
\frac{q_{\mu}q_{\nu}}{m_n^2}\Bigg]
\label{Unitary}~.
\end{equation} 
Notice that, as in the standard model
\begin{equation}
U_{\mu\nu}^{(n)}(q) = \widehat{\Delta}_{\mu\nu}^{(n)}(q,\widehat{a}_n=0)
+ \widehat{\Delta}^{(n)}(q,\widehat{a}_n=0) 
\frac{q_{\mu}q_{\nu}}{m_n^2}~.
\end{equation}
Finally, it is known \cite{dienes:1998vg} that one may reach directly the 
unitary gauge 
by resorting to the following field redefinition
\begin{eqnarray}
A_{\mu}^{(n)} &\rightarrow & A_{\mu}^{(n)} + \partial_{\mu} \theta^{(n)}
\nonumber\\
A_{\chic 5}^{(n)} &\rightarrow & A_{\chic 5}^{(n)} - m_n \theta^{(n)}~,
\end{eqnarray}
which is, at the same time, a non-linear
gauge transformation
with $\theta^{(n)}$ the gauge transformation
parameter and leaves the Lagrangian invariant. The choice 
$\theta^{(n)} = m_n^{-1} A_{\chic 5}^{(n)}$ 
eliminates the scalar component
completely, leading to the unitary propagator of Eq.~(\ref{Unitary})
for all massive vector bosons.  

Given the important differences in the
form of the propagators derived following the
two gauge-fixing procedures presented above,
it is important to verify whether one arrives
at the same answer for physical observables
calculated within either scheme. 
Even though no formal proof to that effect will
be offered here, the next sections are
devoted to the study of the above question in the context
of specific examples.
To accomplish that within a well-defined framework
we will introduce an Abelian interaction at the
level of the five-dimensional Lagrangian, allowing 
photons to interact with fermions.

\section{The Chiral Fermions}
\label{sec:chiral-fermions}

The orbifold compactification adopted in the previous section
allows one to remove from the spectrum the unwanted massless scalar 
corresponding to the zero mode of the fifth component $A_{\chic 5}^{(0)}$
of the photon. As explained in \cite{georgi:2000wb}
this introduces a chiral
structure which appears at the level of the effective four-dimensional
theory, even though the five-dimensional Lagrangian one starts with
is vector-like. In this section we will carry out this construction
in detail, not only for the kinetic term, but also for the interaction
term. The resulting interactions involve four different types of 
vertices, whose Lorentz structure corresponds to vector, axial, scalar
and pseudo-scalar.
We start with the Lagrangian 
\begin{eqnarray}
\cl _{\psi }(x)&=&
\int _{0}^{L}dy\bar{\psi }(x,y)\gamma^{\chic M}
\left( i\partial _{\chic M}+e_{\chic 5} A_{\chic M}\right) \psi (x,y)\nonumber\\
&=&\int ^{L}_{0}dy\bar{\psi }(x,y)
\left( i\db -\gamma_{\chic 5}\partial_y+
e_{\chic 5}\A (x,y)+ie_{\chic 5}\gamma_{\chic 5}A_{\chic 5}(x,y)\right) \psi (x,y)
\label{Lpsi}~,
\end{eqnarray}
with
$\gamma^{\chic M =0,1,2,3}=\gamma^{\mu}$ and
$\gamma^{\chic 4}= i \gamma_{\chic 5}$, where
$\gamma_{\chic 5}$ is defined as usual as
$\gamma_{\chic 5}= i\gamma^0\gamma^1\gamma^2\gamma^3$. We also used the
standard notation $\db=\gamma^\mu\partial_\mu$ and $\A=\gamma^\mu A_\mu$.
Finally, $e_{\chic 5}$ denotes the five-dimensional gauge coupling.

It is elementary to verify that, in the absence of a mass term,
${\cal L}_{\psi}$ is invariant under the transformation
\begin{equation}
\psi(x,y) \rightarrow e^{i\alpha} \gamma_{\chic 5} \psi(x,C-y)\, ,\ \ 
A_\mu(x,y) \rightarrow A_\mu(x,C-y)\, ,\ \
A_{\chic 5}(x,y) \rightarrow -A_{\chic 5}(x,C-y)
\end{equation}
where $\alpha$ is arbitrary and $C=L$ in order to
map the interval  $y \in [0, L=\pi R ]$ into itself.
If we require that
\begin{equation}
\psi \rightarrow \psi' \rightarrow \psi'' = \psi~,  
\end{equation}
we have that  $e^{i\alpha} = \pm 1$.
Next we should impose boundary conditions, i.e. periodicity properties
outside the interval $[0,L]$. One can impose
\begin{equation}
\psi(x,y) = \psi'(x,L+y) = \pm\gamma_{\chic 5} \psi(x,-y)
\label{eq:bc}
\end{equation}
for every $y$ (inside or outside the interval $[0,L]$).
Next one requires that the above relation be satisfied also for the 
transformed fields, i.e.
\begin{equation}
\pm \gamma_{\chic 5}\psi(x,L-y)= 
\pm \gamma_{\chic 5}( \pm \gamma_{\chic 5}\psi(x,L+y) = \psi(x,L+y)~. 
\end{equation}
Since this last equation  
 is satisfied for every $y$, it is also satisfied for
  $y= L+y$,  from which follows that
\begin{equation}
\psi(x,y)=\psi(x,y+2L)~.
\end{equation}
Then we carry out
the Fourier expansion for the $\psi$ field in the interval
$[-L,L]$ (or equivalently expand the even and odd components in
Fourier-cosine and Fourier-sine series, respectively, 
in the interval $[0,L]$), i.e.
\begin{equation}
\psi(x,y) =\frac{1}{\sqrt{L}}\Bigg[
\psi_{\chic R}^{(0)} + \sqrt{2} \sum_{n=1}^{\infty} 
\bigg(\psi_{\chic R}^{(n)}(x) \cos (m_n y)
+\psi_{\chic L}^{(n)}(x) \sin (m_n y)\bigg)\Bigg]~. 
\end{equation}
Imposing the boundary condition of Eq.~(\ref{eq:bc}) with the $+$ sign, 
we see that
\begin{equation}
\gamma_{\chic 5}\psi_{\chic R}^{(n)}(x) =\psi_{\chic R}^{(n)}(x) ,\, \,\,
\gamma_{\chic 5}\psi_{\chic L}^{(n)}(x) =-\psi_{\chic L}^{(n)}(x) ~,
\end{equation}
i.e. $\psi_{\chic R}^{(n)}(x)$ and $\psi_{\chic L}^{(n)}(x)$
can be identified with the chiral right-handed 
and left-handed components, respectively. 

Next we carry out the KK construction 
and we see that all the modes with $n>0$ pick up a mass,
whereas the zero mode ($n=0$) remains massless, i.e.
\begin{equation}
{\cal L}_{\psi,{\chic Q}}= 
\bar{\psi}_{\chic R}^{(0)} 
(i\gamma_{\mu} \partial^{\mu})\psi_{\chic R}^{(0)}
+  \sum_{n=1}^{\infty} \bar{\psi}^{(n)}
(i\gamma_{\mu} \partial^{\mu}+m_n )\psi^{(n)}~,
\end{equation}
where $\psi^{(n)}=\psi_{\chic R}^{(n)}+\psi_{\chic L}^{(n)}$

Using the Fourier decompositions given above and the
auxiliary formulas (notice that since the five-dimensional Lagrangian is
even under a reflection in the fifth component we can change
$\int_0^L dy \rightarrow 1/2 \int_{-L}^L dy$)
\begin{eqnarray} 
\frac{2}{L} \int_{-L}^{L}dy
\cos (m_i y)
\cos (m_j y)
\cos (m_k y)
&=& \delta_{i,j+k} + \delta_{k,i+j} + \delta_{j,i+k}
\nonumber\\
\frac{2}{L} \int_{-L}^{L}dy
\sin (m_i y)
\sin (m_j y)
\cos (m_k y)
&=& 
\delta_{i,j+k} - \delta_{k,i+j} + \delta_{j,i+k}
   ~,
\end{eqnarray}
where $i,j,k\geq 1$, we obtain the following expressions for 
the interacting part ${\cal L}_{\psi,{\chic I}}$ of the Lagrangian 
\begin{equation}
{\cal L}_{\psi,{\chic I}} 
= {\cal L}_{\psi,{\chic I}}^{(0)}+
{\cal L}_{\psi,{\chic I}}^{(0K)}+
{\cal L}_{\psi,{\chic I}}^{(K)}~,
\end{equation}
with
\begin{eqnarray}
\cl^{(0)}_{\psi,\chic I}(x)&=& 
e\bar{\psi}_{\chic R}^{(0)}\A^{(0)} {\psi}_{\chic R}^{(0)} \nonumber\\
\cl ^{(0K)}_{\psi,\chic I}(x) &=&
e\sum_{n=1}\bar{\psi}^{(n)}\A^{(0)} \psi ^{(n)}+
e\sum_{n=1}
\left[\left(\bar{\psi}^{(0)}_{\chic R} \A ^{(n)}\psi^{(n)}_{\chic R} 
+{\mathrm h.c.}\right) +\left(-i \bar{\psi}^{(0)}_{\chic R}A^{(n)}_{\chic 5}
\psi^{(n)}_{\chic L}+{\mathrm h.c.}\right)\right]
\nonumber\\
\cl^{(K)}_{\psi,\chic I}(x) &=&\frac{e}{\sqrt{2}}\sum_{m,n}
\left( \bar{\psi }^{(n+m)}\left( \A ^{(m)}-iA^{(m)}_{\chic 5}\right) 
\psi^{(n)}+{\mathrm h.c.}\right)\nonumber\\ 
&+&\frac{e}{\sqrt{2}}\sum_{m,n}
\bar{\psi }^{(m)}
\left( \A^{(n+m)}+iA^{(n+m)}_{\chic 5}\right) 
\gamma _{\chic 5}\psi^{(n)}
\, ,
\end{eqnarray}
where we have rewritten the five-dimensional coupling, $e_5$ in terms of the
four-dimensional coupling $e$ as $e\rightarrow e_{\chic 5}\sqrt{L}$.

As we can see from the above Lagrangian, 
there are four types of interaction vertices, shown in Fig.~3, with 
different Lorentz structures: $\Gamma_{V}^{\mu}$
is vector, 
$\Gamma_{A}^{\mu}$ is axial, $\Gamma_{S}$ is scalar,
and $\Gamma_{P}$ pseudo-scalar. Notice the conservation of
the KK number at each elementary vertex. 
One direct field-theoretical consequence of this
conservation, in addition to those already mentioned in
the introduction, is that there is no 
higher order mixing between the different bosonic KK modes.
Therefore the gauge boson mass-eigenstates
derived in the previous section do not get modified 
by quantum corrections. 

\section{The gauge invariance of the tree-level $S$-matrix}
\label{sec:gauge-invariance}

In the last two sections we have derived the tree-level
expressions for the vector bosons and scalar fields appearing
in the Lagrangian, as well as their
interaction vertices with the fermions. As a basic application
and a useful self-consistency check we will next 
demonstrate explicitly the gauge invariance
(independence of the gauge-fixing parameter $a$) of the tree-level
$S$-matrix . In addition, we will show that one arrives at the
same expressions for the $S$-matrix regardless of
whether one uses the expressions
for the propagators obtained before or after the compactification
(viz. Eq.~(\ref{Props}) and Eq.~(\ref{Unitary}))
We will show how the cancellations proceed using 
special examples and carrying out the explicit calculation;
a more formal proof will be presented in section~\ref{sec:massive-modes}.
We will study two different types of scattering processes,
one that involves neutral fermions as external particles,
 and one that involves charged ones.

Let us first do the neutral case:
The tree-level
scattering amplitude
${\cal S_{N}}$
for the process
$\bar{\psi}^{(n)}\psi^{(n)} \to \bar{\psi}^{(n)}\psi^{(n)}$ involving the
$n$-th KK fermion of mass $m_n$ is given by (Fig.~4) (we consider only
$s$-channel amplitudes; the argument for $t$-channel amplitudes is identical)
\begin{equation}
{\cal S_{N}}=
\Gamma_{A}^{\mu\, (2n)} 
\Delta_{\mu\nu}^{(n)}(q) \Gamma_{A}^{\nu\, (2n)} 
+ \Gamma_{A}^{\mu\, (2n)}\Delta_{\mu}^{(2n)}(q) \Gamma_{P}^{(2n)}
+ \Gamma_{P}^{(2n)}\Delta_{\nu}^{(2n)}(q)\Gamma_{A}^{\nu\, (2n)} 
+ \Gamma_{P}^{(2n)} \Delta^{(2n)}(q) \Gamma_{P}^{(2n)}~,
\end{equation}
where the elementary vertices $\Gamma_{A}$, etc are given in Fig.~3.
Using that
\begin{equation}
q_{\mu} \Gamma_{A}^{\mu\, (2n)} = -2 i m_n \Gamma_{P}^{(2n)}  
\label{Id1}
\end{equation}
we see that the condition for the cancellation of the
terms with the double poles, proportional to $(1-a)$, is simply
\begin{equation}
4 m_n^2 - 4 m_n m_{2n} + m_{2n}^2 = 0   ~,
\label{Cond}
\end{equation}
which is automatically satisfied by virtue of Eq.~(\ref{Mass}). 
Notice that the omission of the mixing terms would result in
a residual dependence of the $S$ matrix on $a$. 
After the double poles have canceled, the remaining 
contribution is effectively given
by fixing $a=1$. It is elementary
to verify that the result obtained in that case
is identical to the one computed  
by using the unitary-type of gauge given 
in Eq.~(\ref{Unitary}), derived when the gauge fixing is carried 
out after the compactification; 
indeed, the two answers
coincide provided that $ (2m_n)^2/m_{2n}^2 =1$, which 
is the same condition given in Eq.~(\ref{Cond}), and 
holds for every $n$ . 

Turning to the charged case, it is straightforward to see that
again due to the special mass relations the dependence
on  $a$ cancels, and the final answer coincides
with the Feynman and unitary gauges.
For example, the amplitude ${\cal S_{C}}$
for the process 
$\bar{\psi}^{(\ell)}\psi^{(i+j)} \to \bar{\psi}^{(i)}\psi^{(\ell+j)}$
is given by
\begin{equation}
{\cal S_{C}}=
\Gamma_{V}^{\mu\, (j)} 
\Delta_{\mu\nu}^{(j)}(q) \Gamma_{V}^{\nu\, (j)} 
+ \Gamma_{A}^{\mu\, (j)}\Delta_{\mu}^{(j)}(q) \Gamma_{S}^{(j)}
+ \Gamma_{S}^{(j)}\Delta_{\nu}^{(j)}(q)\Gamma_{V}^{\nu\, (j)} 
+ \Gamma_{S}^{(j)} \Delta^{(j)}(q) \Gamma_{S}^{(j)}~.
\end{equation}
Using that
\begin{equation}
q_{\mu} \Gamma_{V}^{\mu\, (j)} = i (m_{i+j}-m_i) \Gamma_{S}^{(j)}  
\label{Id2}~,
\end{equation}
we find that the condition for the gauge cancellation reads
\begin{equation}
(m_{i+j}-m_i)^{2} - m_j (m_{i+j}-m_i) -m_j (m_{\ell +j}-m_{\ell})
+  m_j^2 = 0~,
\end{equation}
which is again automatically satisfied, since  $m_{i+j}-m_i =m_j$. 

\section{Giving mass to the zero modes}
\label{sec:massive-modes}

In order to  recover conventional QED with massive  electrons from the
5-d construction presented in the  previous section, we must give mass
to the  fermionic zero modes.  In  this section we will  carry out the
construction which gives  masses to the zero modes.  The presence of a
new mass-term leads to  the need of redefining the mass-eigenstates of
the theory, which in turn alters the form of the interaction term.

The construction we will adopt for giving
mass to the fermionic zero modes proceeds as follows (for an alternative
construction see \cite{dvali:2000ha}).
We introduce an additional set of fermions, to be denoted  
by $\chi$, which have the opposite chirality properties 
compared to the $\psi$. Specifically, add to the Lagrangian density
${\cal L}_{\psi}$
of Eq.~(\ref{Lpsi}) the  Lagrangian density ${\cal L}_{\chi}$
of a new fermionic field $\chi$,
obtained from ${\cal L}_{\psi}$ by simply changing   
$\psi\to \chi$,
and impose the following transformation properties on the
$\psi$ and $\chi$ fields
\begin{eqnarray}
\psi(x,y) & \rightarrow & \psi' = \, \gamma_{\chic 5}\psi(x,L-y) \nonumber\\ 
\chi(x,y) & \rightarrow & \chi' = -\gamma_{\chic 5}\psi(x,L-y) ~.
\end{eqnarray}
Then, a mass-term is allowed, which does not violate the
chiral symmetry, 
\begin{equation}
{\cal L}_{\mathrm mass} = -\int^{L}_{0}dy m_{0}\left( \bar{\psi }(x,y)\chi (x,y)+\bar{\chi }(x,y)\psi (x,y)\right)
\end{equation}
giving rise to the final Lagrangian
\begin{equation}
{\cal L} = {\cal L}_{\psi} +  {\cal L}_{\chi} + {\cal L}_{mass}~.
\end{equation}
The fact that $\chi$ has the opposite 
chiral transformation property
than $\psi$ results in the following Fourier expansion 
\begin{equation}
\chi(x,y) = \frac{1}{\sqrt{L}}\Bigg[
\chi_{\chic L}^{(0)}(x) + \sqrt{2} \sum_{n=1}^{\infty} 
\bigg(\chi_{\chic L}^{(n)}(x) \cos (m_n y)
+\chi_{\chic R}^{(n)}(x) \sin (m_n y)\bigg) \Bigg] ~,
\end{equation}
and after the standard KK construction 
we find for the quadratic part of the Lagrangian
\begin{equation}
\cl _{Q}(x)=\bar{f}^{(0)}\left( i\db -m_{0}\right) f^{(0)}+
\sum _{n=1}\bar{\xi }^{(n)}i\db \xi ^{(n)}-\bar{\xi }^{(n)}M_{n}\xi ^{(n)}~,
\end{equation}
where 
\begin{equation}
f^{(0)}=\psi _{\chic R}^{(0)}+\chi _{\chic L}^{(0)},
\qquad \xi ^{(n)}=\left( \begin{array}{c}
\psi ^{(n)}_{\chic L}+\psi ^{(n)}_{\chic R}\\
\chi ^{(n)}_{\chic L}+\chi ^{(n)}_{\chic R}
\end{array}\right) ,\qquad M_{n}=\left( \begin{array}{cc}
-m_{n} & m_{0}\\
m_{0} & m_{n}
\end{array}\right) ~.
\end{equation}
The interaction terms can be separated 
as before into three pieces, one containing only zero modes,
one containing a mixture of zero modes
and KK modes, and one containing only KK modes, i.e.
\begin{equation}
{\cal L}_{{\chic I}} 
= {\cal L}_{{\chic I}}^{(0)}+
{\cal L}_{{\chic I}}^{(0K)}+
{\cal L}_{{\chic I}}^{(K)}~,
\end{equation}
with  
\begin{eqnarray}
\cl ^{(0)}_{\chic I}(x)&=& e\bar{f}^{(0)}\A^{(0)} f^{(0)} \nonumber\\
\cl ^{(0K)}_{\chic I}(x) &=&
e\sum _{n=1}\bar{\xi }^{(n)}\A^{(0)} \xi ^{(n)}\nonumber \\
&+&
e\sum _{n=1}\left[
\left( \bar{f}^{(0)}\A ^{(n)}
\left( \psi ^{(n)}_{\chic R}+\chi ^{(n)}_{\chic L}\right) 
+\mathrm{h}.c.\right) +\left( -i\bar{f}^{(0)}A_{\chic 5}^{(n)}
\left( \psi ^{(n)}_{\chic L}-\chi ^{(n)}_{\chic R}\right)
+\mathrm{h}.c.\right) \right]
\nonumber\\
\cl ^{(K)}_{\chic I}(x)&=&\frac{e}{\sqrt{2}}\sum _{m,n}
\left( \bar{\xi }^{(n+m)}\left( \A ^{(m)}-
i\sigma_{\chic 3}A^{(m)}_{\chic 5}\right) 
\xi ^{(n)}+{\mathrm h.c.}\right) \nonumber \\
&+&\frac{e}{\sqrt{2}}\sum _{m,n}
\bar{\xi }^{(m)}
\left( \A ^{(n+m)}\sigma_{\chic 3}+iA^{(n+m)}_{\chic 5}\right) \gamma_{\chic 5}\xi ^{(n)}
   ~,
\end{eqnarray}
where \( \sigma_{\chic 3} \) is the Pauli matrix and acts on the fields \( \xi ^{(n)}. \)
Next one 
should rewrite the interaction Lagrangian in terms of the mass eigenstates;
this may be done in different ways. 
The mass matrix \( M_{n} \) is real and
symmetric, so it can be diagonalized by an orthogonal transformation. Since
the corresponding
eigenvalues are \( \pm \lambda _{n}\equiv \pm \sqrt{m^{2}_{n}+m^{2}_{0}} \)
we have 
\begin{equation}
U_{n}^{\dagger }M_{n}U_{n}=-\lambda _{n}\sigma_{\chic 3},\qquad U_{n}=\left( \begin{array}{cc}
c_{n} & s_{n}\\
-s_{n} & c_{n}
\end{array}\right) ,\qquad c_{n}=
\sqrt{\frac{\lambda _{n}+m_{n}}{2\lambda_{n}}},\qquad
s_{n}=\sqrt{\frac{\lambda _{n}-m_{n}}{2\lambda _{n}}}~,
\end{equation}
which obviously satisfy the relations
\begin{equation}
c^{2}_{n}+s^{2}_{n}=1,\qquad c^{2}_{n}-s^{2}_{n}=\frac{m_{n}}{\lambda
_{n}},\qquad 2c_{n}s_{n}=\frac{m_{0}}{\lambda _{n}}\, .
\end{equation}
Then we can use the matrix $U_n$ to diagonalize the quadratic 
 term by defining a new set of fields $\xi ^{(n)}$ and $f^{(n)}$ 
as follows
\begin{equation}
\xi ^{(n)}=U_n f^{(n)},\qquad f^{(n)}=\left( \begin{array}{c}
f_{1}^{(n)}\\
f_{2}^{(n)}
\end{array}\right)\, .
\label{fieldred}
\end{equation}
In terms of these fields
the quadratic term is rewritten as 
\begin{equation}
\label{eq:kinetic}
\cl _{\chic Q}(x)=\bar{f}^{(0)}\left( i\db -m_{0}\right) f^{(0)}
+\sum _{n=1}\bar{f}^{(n)}i\db f^{(n)}+\lambda _{n}\bar{f}^{(n)}
\sigma_{\chic 3}f^{(n)}\, .  
\label{fieldredkin}
\end{equation}
Notice that the fields \( f^{(n)}_{2} \) have mass \( \lambda _{n} \) while
the fields \( f^{(n)}_{1} \) have mass \( -\lambda _{n} \). This sign can
be reversed by further 
redefining \( f^{(n)}_{1}\rightarrow \gamma _{\chic 5}f^{(n)}_{1} \);
in that case we would
obtain two completely degenerate Dirac fields.
 For the time
being however we will continue working with the basis 
in which the two fermions have masses
with opposite sign. 
The field redefinition of Eq.~(\ref{fieldred})
immediately leads to the following interaction
Lagrangians
\begin{eqnarray}
\cl ^{(0)}_{\chic I}(x) &=& e\bar{f}^{(0)}\A^{(0)} f^{(0)} \nonumber\\
\cl ^{(0K)}_{\chic I}(x) &=&
e\sum _{n=1}\bar{f}^{(n)}\A^{(0)} f^{(n)}\nonumber\\
&+&
e\sum _{n=1}\left( \bar{f}\A ^{(n)}\left(
c_{n}f^{(n)}_{1\chic R}+s_{n}f^{(n)}_{2\chic R}-s_{n}f^{(n)}_{1\chic L}+
c_{n}f^{(n)}_{2\chic L}\right) +\mathrm{h}.c.\right) 
\nonumber\\
&+& e\sum _{n=1}\left( -i\bar{f}A_{\chic 5}^{(n)}
\left( c_{n}f^{(n)}_{1\chic L}+s_{n}f^{(n)}_{2\chic L}+s_{n}f^{(n)}_{1\chic R}-c_{n}f^{(n)}_{2\chic R}\right) +\mathrm{h}.c.\right) \nonumber\\
\cl ^{(K)}_{\chic I}(x)&=&
\frac{e}{\sqrt{2}}\sum _{m,n}
\left( \bar{f}^{(n+m)}U^{\dagger }_{n+m}
\left( \A ^{(m)}-i\sigma_{\chic 3}A^{(m)}_{\chic 5}\right) 
U_{n}f^{(n)}+\mathrm{h}.c.\right)\nonumber\\ 
&+& \frac{e}{\sqrt{2}}\sum _{m,n}\bar{f}^{(m)} U_{m}^{\dagger }
\left( \A ^{(n+m)}\sigma_{\chic 3}+iA^{(n+m)}_{\chic 5}\right) \gamma _{\chic 5} U_{n}f^{(n)}
\label{fieldredint}~.
\end{eqnarray}
Next we turn to the issue of the 
gauge-invariance of the tree-level
$S$-matrix, and the current (non)conservation relations 
which enforce it. In particular, we will derive
the Ward identities relating the various terms in the
interaction Lagrangian of Eq.~(\ref{fieldredint}); they
will constitute the generalizations of the elementary tree-level 
Ward identities employed in section~\ref{sec:gauge-invariance}, i.e. 
of Eq.~(\ref{Id1}) and  Eq.~(\ref{Id2}).

From the kinetic terms  of Eq.~(\ref{fieldredkin})
we obtain the following
Dirac equations:
\begin{equation}
\db f^{(0)}=-im_{0}f^{(0)},\qquad \bar{f}^{(0)}\overleftarrow{\db }=im_{0}\bar{f}^{(0)},\qquad \db f^{(n)}=i\lambda _{n}\sigma_{\chic 3}f^{(n)},\qquad \bar{f}^{(n)}\overleftarrow{\db }=-i\lambda _{n}\bar{f}^{(n)}\sigma_{\chic 3}~.
\end{equation}
By employing these Dirac equations
one can compute the following divergence
\begin{eqnarray}
\partial _{\mu }\left( \bar{f}^{(m)}
U_{m}^{\dagger }\sigma_{\chic 3}U_{n}\gamma ^{\mu }
\gamma _{\chic 5}f^{(n)}\right) &=&\bar{f}^{(m)}\overleftarrow{\db }
U_{m}^{\dagger }\sigma_{\chic 3}U_{n}
\gamma _{\chic 5}f^{(n)}-\bar{f}^{(m)}U_{m}^{\dagger }
\sigma_{\chic 3}U_{n}\gamma _{\chic 5}\db f^{(n)}\nonumber\\
&=&-i\bar{f}^{(m)}\lambda _{m}\sigma_{\chic 3}U_{m}^{\dagger }\sigma_{\chic 3}U_{n}\gamma _{\chic 5}f^{(n)}-i\bar{f}^{(m)}U_{m}^{\dagger }\sigma_{\chic 3}U_{n}\sigma_{\chic 3}\lambda _{n}\gamma _{\chic 5}f^{(n)}
\nonumber\\
&=&i\bar{f}^{(m)}\left( U_{m}^{\dagger }M_{m}\sigma_{\chic 3}U_{n}+U_{m}^{\dagger }\sigma_{\chic 3}M_{n}U_{n}\right) \gamma _{\chic 5}f^{(n)}\nonumber\\
&=&i\bar{f}^{(m)}\left( U_{m}^{\dagger }\left( M_{m}\sigma_{\chic 3}+\sigma_{\chic 3}M_{n}\right) U_{n}\right) \gamma _{\chic 5}f^{(n)}\nonumber\\
&=&-i(m_{m}+m_{n})\bar{f}^{(m)}\left( U_{m}^{\dagger }
U_{n}\right) \gamma _{\chic 5}f^{(n)}
\label{CE1}~.
\end{eqnarray}
In deriving Eq.~(\ref{CE1}) 
we have used the relations
\begin{equation}
M_{n}U_{n}=-U_{n}\sigma_{\chic 3}\lambda _{n},\qquad U_{m}^{\dagger
}M_{m}=-\lambda _{m}\sigma_{\chic 3}U^{\dagger }_{m}~,
\end{equation}
which follow directly from the fact that 
\( U_{n}^{\dagger }M_{n}U_{n}=-\lambda _{n}\sigma_{\chic 3} \). 
 
The identity of Eq.~(\ref{CE1}) guarantees the correct 
gauge cancellations for $S$-matrix elements involving
these interaction terms. Notice that 
for the special case $m=n$ Eq.~(\ref{CE1}) reduces to Eq.~(\ref{Id1})

Similarly, for the other two interacting terms in 
Eq.~(\ref{fieldredint}) we have
\begin{eqnarray}
\partial _{\mu }\left( \bar{f}^{(m+n)}U_{m+n}^{\dagger }U_{n}\gamma ^{\mu }f^{(n)}\right) &=&\bar{f}^{(m+n)}\overleftarrow{\db }U_{m+n}^{\dagger }U_{n}f^{(n)}+
\bar{f}^{(m+n)}U_{m+n}^{\dagger }U_{n}\db f^{(n)}\nonumber\\
&=&-i\bar{f}^{(m+n)}\lambda _{m+n}\sigma_{\chic 3}U_{m+n}^{\dagger }U_{n}f^{(n)}+i\bar{f}^{(m+n)}U_{m+n}^{\dagger }U_{n}\sigma_{\chic 3}\lambda _{n}f^{(n)}\nonumber\\
&=&i\bar{f}^{(m+n)}\left( U_{m+n}^{\dagger }M_{m+n}U_{n}
-U_{m+n}^{\dagger }M_{n}U_{n}\right) f^{(n)} \nonumber\\
&=& i\bar{f}^{(m+n)}U_{m+n}^{\dagger }\left( M_{m+n}-M_{n}\right) U_{n}f^{(n)}
\nonumber\\
&=&-im_{m}\bar{f}^{(m+n)}U_{m+n}^{\dagger }\sigma_{\chic 3}U_{n}f^{(n)}
\label{CE2}~,
\end{eqnarray}
which again guarantees gauge invariance of diagrams including these couplings
Again, Eq.~(\ref{CE2}) is the generalization of  Eq.~(\ref{Id2})
Similar arguments apply
to the couplings involving one zero mode fermion.

The diagonalization presented above 
leads to
a rather compact Lagrangian 
(except perhaps for the terms containing zero modes). On the other hand,
the fact that $f^{(n)}_1$ and $f^{(n)}_2$ have masses with different signs
may appear unappealing; even though this 
may be remedied by means of the the field-redefinition mentioned earlier
it would spoil the simple structure of the 
interaction Lagrangian .
In addition,
it would be 
difficult to exploit the degeneracy (once the wrong
sign in the mass is removed by a chiral transformation) of the pair of Dirac
fermions. We now present an alternative diagonalization which could be useful
in some cases. We write quadratic terms for the KK modes in the following form 
\begin{equation}
\cl _{\chic Q}=\sum _{n=1}\bar{\xi }_{\chic R}^{(n)}
i\db \xi _{\chic R}^{(n)}+\bar{\xi }_{\chic L}^{(n)}i\db \xi _{\chic
L}^{(n)}-\bar{\xi }_{\chic R}^{(n)}M_{n}\xi _{\chic L}^{(n)}-\bar{\xi }_{\chic L}^{(n)}M_{n}\xi _{\chic R}^{(n)}~.
\end{equation}
Now since we have that 
\begin{equation}
M^{2}_{n}=\lambda ^{2}_{n}\end{equation}
we find that the matrix \( M_{n}/\lambda _{n} \) is unitary, symmetric, 
self-adjoint
and real, so we can define
\begin{equation}
g^{(n)}_{\chic R}=\xi _{\chic R}^{(n)},\qquad g_{\chic
L}^{(n)}=\frac{M_{n}}{\lambda _{n}}\xi _{\chic L}^{(n)}~,
\end{equation}
and find 
\begin{eqnarray}
\cl _{\chic Q }&=&\sum _{n=1}\bigg[
\bar{g}_{\chic R}^{(n)}i\db g_{\chic R}^{(n)}+\bar{g}_{\chic L}^{(n)}i\db g_{\chic L}^{(n)}-\lambda _{n}\left( \bar{g}_{\chic R}^{(n)}g_{\chic L}^{(n)}+\bar{g}_{\chic L}^{(n)}g_{\chic R}^{(n)}\right)\bigg]
\nonumber\\ 
&=& \sum _{n=1}\bigg[\bar{g}^{(n)}i\db g^{(n)}-
\lambda _{n}\bar{g}^{(n)}g^{(n)}\bigg]~.
\end{eqnarray}
In addition this Lagrangian is invariant under any rotation of the \( g \)
fields (\( g_{\chic L} \) and \( g_{\chic R} \) must be rotated in the same way). 
Although this
formulation 
is rather compact and leads naturally to positive masses it is not obvious
how to simplify further the interaction Lagrangian, 
since it treats differently left
from right components. The two basis are related by the following
transformation
\begin{equation}
g^{(n)}_{\chic R}=U_{n}f_{\chic R}^{(n)},\qquad 
g_{\chic L}^{(n)}=\frac{M_{n}}{\lambda _{n}} U_n f_{\chic
L}^{(n)}=-U_{n}\sigma_{\chic 3}f_{\chic L}^{(n)}~.
\end{equation}

\section{Conclusions}
\label{sec:conclusions}

We have constructed a generalization of QED in five dimensions with
the extra dimension
compactified on a $S^1/{\mathbb Z}_2$ orbifold. This type of 
compactification 
leads naturally to a low energy theory with just one massless Weyl 
fermion and one Abelian gauge boson. 
We have derived the Feynman rules of four-dimensional theory 
obtained after
compactification which contain an infinity of Kaluza-Klein modes.
In particular we have focused on the issue of the gauge fixing, and 
have derived the tree-level propagators when the gauge fixing 
has been carried out before or after the compactification of
the extra dimension. It turns out
that in the former case the derivation of tree-level boson propagators
is rather involved, at least in the context of the
covariant (in 5-d) linear 
gauge fixing enforced by the addition of a term 
$\frac{1}{2a} (\partial_{\chic M} A^{\chic M})^2$ to the original
Lagrangian.
The  resulting expressions for the propagators have rather particular
features; for  example, one cannot  arrive at a unitary-gauge  type of
spectrum for any choice of the gauge-fixing parameter $a$, 
nor can one choose $a$ is such a way
as to give rise to  transverse, Landau gauge type of propagators.  The
subtleties associated with the gauge-fixing in this case can be traced
back to the fact that the standard  KK construction introduces at
the level of the 4-d  effective Lagrangian a residual tree-level mixing
between  the   vector  bosons  (photons)  and   their  (scalar)  fifth
components;  the  latter must  be  properly  taken  into account  when
deriving the  various tree-level propagators.   On the other  hand, if
the gauge fixing is carried out after the compactification one arrives
at  a  Standard  Model-like  result  for the  various  propagators;  in
particular  both the  unitary  and the  Landau gauges  can be  reached
continuously. The Landau gauge defined after compactification can be 
useful when computing the effective potential in theories with
extra dimensions. Notice however that  in that  case graphs  with 
additional  massless scalars  
need be considered, which in general are non-vanishing.
Notice in addition that 
(a) both constructions give rise to the same
propagators in the Feynman gauge, and (b) the unitary gauge
may be obtained directly before compactification
by means of a non-covariant, axial-gauge type of gauge fixing. 
In addition, we have demonstrated how the gauge cancellations
proceed in tree-level $S$ matrix elements, and that the 
two gauge-fixing procedures (before and after the compactification)
give rise to the same answer. 

Four-dimensional QED is vector-like and involves massive 
Dirac fermions, however the orbifold compactification 
we have described above starting with
only one fermion leads, at low energies, to only one massless 
chiral fermion. To obtain QED as a low-energy theory we have
included an additional five-dimensional fermion in the spectrum 
which has opposite transformation properties with respect to the 
orbifold symmetry. Then, one can write a mass term which preserves the
orbifold symmetry. The low-energy limit of this theory is a
theory with just one Abelian gauge boson and one massive Dirac fermion.
The spectrum of fermionic KK modes is doubled and displays
non-trivial mixing among the different components at each KK level. 
This requires a simple diagonalization which, however, complicates
the structure of the interactions.
In this theory we have verified, by means of compact identities 
involving the divergences of currents, that the gauge cancellations 
demonstrated before go through after the final diagonalization.

The above considerations provide a well-defined 
minimal calculational framework which allows for the further 
detailed study of theories with one universal extra dimension.

\acknowledgements

This work has been funded by CICYT under the Grant AEN-99-0692, by DGESIC under
the Grant PB97-1261 and by the DGEUI of the ``Generalitat Valenciana'' 
under the Grant GV98-01-80.

\vspace{0.5cm}
\centerline{\large FIGURE CAPTIONS}
\vspace{0.3cm}

\noindent
Fig.1: The three basic tree-level propagators arising after  
covariant 5-d gauge-fixing.

\medskip
\noindent
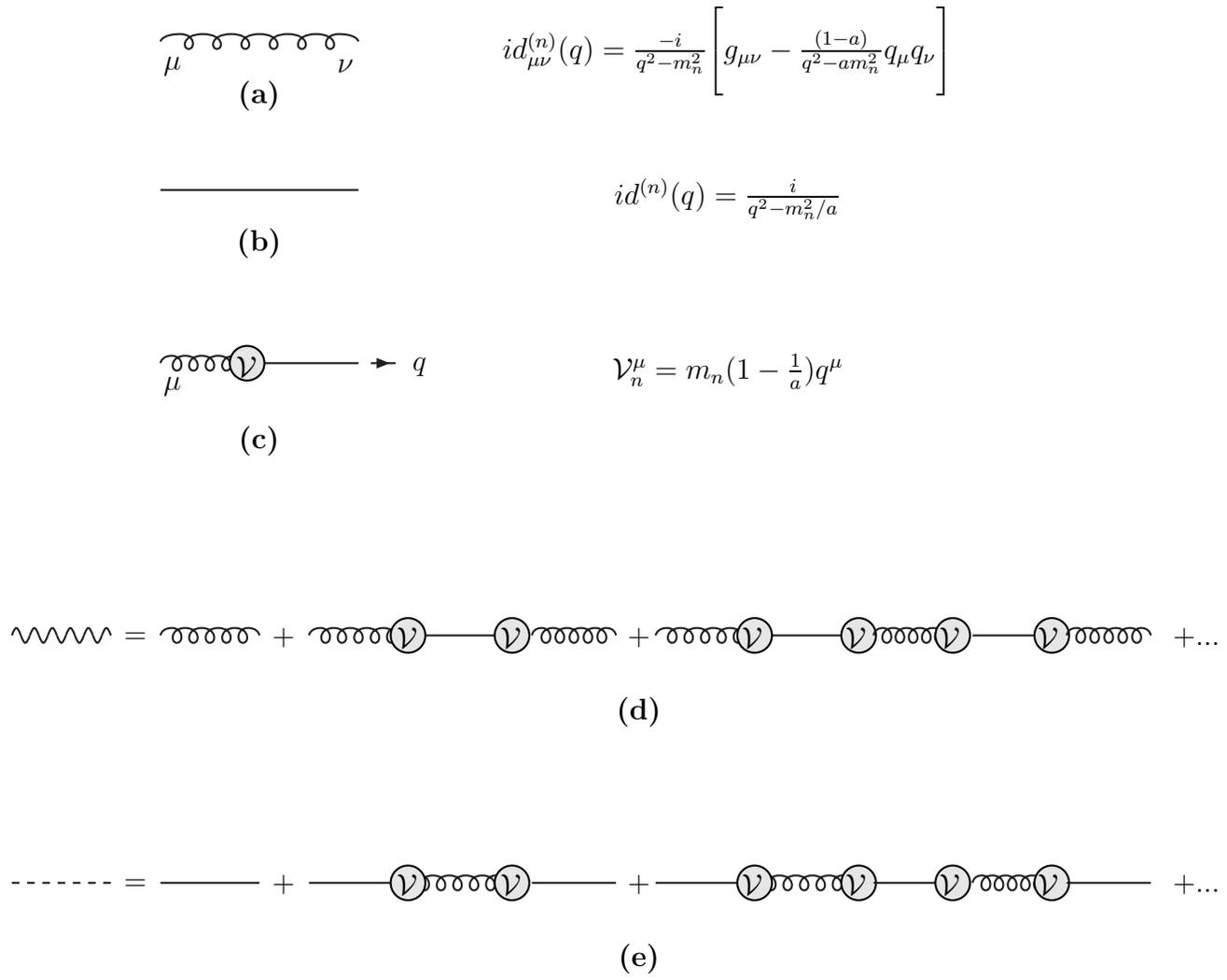
Fig.2: The tree-level propagators arising by ignoring the mixing
term, and the Dyson series obtained by treating the
mixing term as an interaction. 

\medskip
\noindent
Fig.3: The four types of interaction vertices for non-zero modes.

\medskip
\noindent
Fig.4: The tree-level $S$-matrix for the process
$\bar{\psi}^{(n)}\psi^{(n)} \to \bar{\psi}^{(n)}\psi^{(n)}$ .

\medskip
\noindent
Fig.5: The tree-level $S$-matrix for  the process
$\bar{\psi}^{(\ell)}\psi^{(i+j)} \to \bar{\psi}^{(i)}\psi^{(\ell+j)}$.


\begin{center}
\begin{picture}(400,500)(0,0)
\SetWidth{0.8}

\Photon(20,490)(100,490){2}{4.5}
\Text(25,482)[]{$\mu$}
\Text(95,482)[]{$\nu$}
\Text(250,488)[]{$i\Delta_{\mu\nu}^{(n)}(q) =  
-i\Bigg[\frac{g_{\mu\nu}}{q^2-m_n^2}  
- \frac{(1-a)q_{\mu}q_{\nu} }{(q^2-m_n^2)^2}\Bigg]$}
\Text(60,470)[]{\bf (a)}

\DashLine(20,390)(100,390){2}
\Text(250,388)[]{$
i\Delta^{(n)}(q) = i\Bigg[\frac{1}{q^2-m_n^2} + 
\frac{m_n^2(1-a)}{(q^2-m_n^2)^2}\Bigg]$}
\Text(60,370)[]{\bf (b)}

\DashLine(20,290)(60,290){2}
\Photon(60,290)(100,290){2}{3.5}
\Text(95,282)[]{$\nu$}
\Text(250,288)[]{$ 
i\Delta_{\nu}^{(n)}(q) =  \frac{m_n(1-a)}{(q^2-m_n^2)^2}q_{\nu} $}
\Text(-5,290)[]{$q$}
\ArrowLine(0,290)(10,290)
\Text(60,270)[]{\bf (c)}

\Text(180,100)[]{\bf Fig.\ 1}

\end{picture}

\end{center}

\newpage
\begin{center}
\begin{picture}(400,500)(0,0)
\SetWidth{0.8}

\Gluon(20,490)(100,490){3}{6.5}
\Text(25,482)[]{$\mu$}
\Text(95,482)[]{$\nu$}
\Text(250,488)[]{$i d_{\mu\nu}^{(n)}(q) =  
\frac{-i}{q^2-m_n^2}\Bigg[g_{\mu\nu}  
- \frac{(1-a)}{q^2-a m_n^2}q_{\mu}q_{\nu}\Bigg]$}
\Text(60,470)[]{\bf (a)}


\Line(20,430)(100,430)
\Text(250,428)[]{$
i d^{(n)}(q) = \frac{i}{q^2- m_n^2/a}$}
\Text(60,410)[]{\bf (b)}

\Gluon(20,360)(60,360){3}{5.5}
\Line(60,360)(100,360)
\GCirc(55,360){7}{0.9}
\Text(25,352)[]{$\mu$}
\Text(55,360)[]{${\cal V} $}
\Text(250,358)[]{$ {\cal V}^{\mu}_n  =  m_n (1-\frac{1}{a})q^{\mu} $}
\ArrowLine(105,360)(115,360)
\Text(125,360)[]{$q$}
\Text(60,330)[]{\bf (c)}
\Photon(-40,250)(0,250){3}{5.5}
\Text(10,250)[]{$=$}
\Gluon(20,250)(60,250){3}{5.5}
\Text(70,250)[]{$+$}
\Gluon(80,250)(120,250){3}{5.5}
\Line(120,250)(160,250)
\GCirc(120,250){7}{0.9}
\Text(120,250)[]{${\cal V} $}
\GCirc(162,250){7}{0.9}
\Text(162,250)[]{${\cal V} $}
\Gluon(170,250)(204,250){3}{5.5}
\Text(214,250)[]{$+$}
\Gluon(220,250)(260,250){3}{5.5}
\Line(260,250)(300,250)
\GCirc(260,250){7}{0.9}
\Text(260,250)[]{${\cal V} $}
\GCirc(302,250){7}{0.9}
\Text(302,250)[]{${\cal V} $}
\Gluon(308,250)(340,250){3}{5.5}
\GCirc(340,250){7}{0.9}
\Text(340,250)[]{${\cal V} $}
\Line(348,250)(380,250)
\GCirc(380,250){7}{0.9}
\Text(380,250)[]{${\cal V} $}
\Gluon(386,250)(420,250){3}{5.5}
\Text(440,250)[]{$+ ...$}
\Text(214,220)[]{\bf (d)}

\DashLine(-40,150)(0,150){3}
\Text(10,150)[]{$=$}
\Line(20,150)(60,150)
\Text(70,150)[]{$+$}
\Line(80,150)(120,150)
\Gluon(120,150)(160,150){3}{5.5}
\GCirc(120,150){7}{0.9}
\Text(120,150)[]{${\cal V} $}
\GCirc(162,150){7}{0.9}
\Text(162,150)[]{${\cal V} $}
\Line(170,150)(204,150)
\Text(214,150)[]{$+$}
\Line(220,150)(260,150)
\Gluon(260,150)(300,150){3}{5.5}
\GCirc(260,150){7}{0.9}
\Text(260,150)[]{${\cal V} $}
\GCirc(302,150){7}{0.9}
\Text(302,150)[]{${\cal V} $}
\Line(308,150)(340,150)
\GCirc(340,150){7}{0.9}
\Text(340,150)[]{${\cal V} $}
\Gluon(348,150)(380,150){3}{5.5}
\GCirc(380,150){7}{0.9}
\Text(380,150)[]{${\cal V} $}
\Line(386,150)(420,150)
\Text(440,150)[]{$+ ...$}
\Text(214,120)[]{\bf (e)}


\Text(180,0)[]{\bf Fig.\ 2}

\end{picture}

\end{center}

\newpage
\begin{center}
\begin{picture}(400,500)(0,0)
\SetWidth{0.8}

\Photon(40,490)(40,451){2}{3.5}
\ArrowLine(10,400)(40,450)
\ArrowLine(40,450)(70,400)
\Text(51,472)[]{$A_{\mu}^{(j)}$}
\Text(10,420)[]{$\psi^{(i)}$}
\Text(76,420)[]{$\bar{\psi}^{(i+j)}$}
\GCirc(40,451){6}{0.9}
\Text(40,451.4)[]{$V$}
\Text(240,451.4)[]
{$\Gamma_{V}^{\mu\, (j)} : 
(\frac{ie}{\sqrt{2}})\, \bar{\psi}^{(i+j)}\gamma^{\mu}\psi^{(i)}$}

\Photon(40,380)(40,341){2}{3.5}
\ArrowLine(10,290)(40,350)
\ArrowLine(40,350)(70,290)
\Text(55,362)[]{$A_{\mu}^{(i+j)}$}
\Text(10,310)[]{$\psi^{(i)}$}
\Text(78,310)[]{$\bar{\psi}^{(j)}$}
\GCirc(40,344){6}{0.9}
\Text(40,344.8)[]{$A$}
\Text(240,343.4)[]
{$\Gamma_{A}^{\mu\, (i+j)} : (\frac{ie}{\sqrt{2}})\, \bar{\psi}^{(j)}
\gamma^{\mu}\gamma_{5}\psi^{(i)}$}

\DashLine(40,270)(40,231){2}
\ArrowLine(10,180)(40,240)
\ArrowLine(40,240)(70,180)
\Text(49,252)[]{$A_{5}^{(j)}$}
\Text(10,200)[]{$\psi^{(i)}$}
\Text(76,200)[]{$\bar{\psi}^{(i+j)}$}
\GCirc(40,233){6}{0.9}
\Text(40,233.4)[]{$S$}
\Text(240,233.4)[]
{$\Gamma_{S}^{(j)} : 
(\frac{e}{\sqrt{2}})\, \bar{\psi}^{(i+j)}\psi^{(i)}$}

\DashLine(40,160)(40,121){2}
\ArrowLine(10,70)(40,130)
\ArrowLine(40,130)(70,70)
\Text(55,142)[]{$A_{5}^{(i+j)}$}
\Text(10,90)[]{$\psi^{(i)}$}
\Text(78,90)[]{$\bar{\psi}^{(j)}$}
\GCirc(40,123){6}{0.9}
\Text(40,123.4)[]{$P$}
\Text(240,123.4)[]
{$\Gamma_{P}^{(i+j)} : 
-(\frac{e}{\sqrt{2}})\, 
\bar{\psi}^{(j)}\gamma_{5}\psi^{(i)}$}

\Text(180,0)[]{\bf Fig.\ 3}

\end{picture}

\end{center}

\newpage

\begin{center}
\begin{picture}(400,500)(0,0)
\SetWidth{0.8}

\Text(-40,460)[]{${\cal S_{N}}$}
\Text(-10,460)[]{$=$}

\ArrowLine(10,520)(40,490)
\ArrowLine(40,490)(70,520)
\Photon(40,490)(40,430){2}{4.5}
\ArrowLine(10,400)(40,430)
\ArrowLine(40,430)(70,400)
\Text(55,462)[]{$\Delta_{\mu\nu}^{(2n)}$}
\Text(10,420)[]{$\psi^{(n)}$}
\Text(72,420)[]{$\bar{\psi}^{(n)}$}
\Text(10,500)[]{$\psi^{(n)}$}
\Text(72,500)[]{$\bar{\psi}^{(n)}$}
\GCirc(40,430){6}{0.9}
\Text(40,430.4)[]{$A$}
\GCirc(40,490){6}{0.9}
\Text(40,490.4)[]{$A$}

\Text(100,460)[]{$+$}

\ArrowLine(130,520)(160,490)
\ArrowLine(160,490)(190,520)
\DashLine(160,490)(160,430){2}
\ArrowLine(130,400)(160,430)
\ArrowLine(160,430)(190,400)
\Text(175,462)[]{$\Delta^{(2n)}$}
\Text(130,420)[]{$\psi^{(n)}$}
\Text(192,420)[]{$\bar{\psi}^{(n)}$}
\Text(130,500)[]{$\psi^{(n)}$}
\Text(192,500)[]{$\bar{\psi}^{(n)}$}
\GCirc(160,430){6}{0.9}
\Text(160,430.4)[]{$P$}
\GCirc(160,490){6}{0.9}
\Text(160,490.4)[]{$P$}

\Text(220,460)[]{$+$}

\ArrowLine(250,520)(280,490)
\ArrowLine(280,490)(310,520)
\Photon(280,490)(280,460){2}{4.5}
\DashLine(280,460)(280,430){2}
\ArrowLine(250,400)(280,430)
\ArrowLine(280,430)(310,400)
\Text(297,462)[]{$\Delta_{\mu}^{(2n)}$}
\Text(250,420)[]{$\psi^{(n)}$}
\Text(312,420)[]{$\bar{\psi}^{(n)}$}
\Text(250,500)[]{$\psi^{(n)}$}
\Text(312,500)[]{$\bar{\psi}^{(n)}$}
\GCirc(280,430){6}{0.9}
\Text(280,430.4)[]{$P$}
\GCirc(280,490){6}{0.9}
\Text(280,490.4)[]{$A$}

\Text(340,460)[]{$+$}

\ArrowLine(370,520)(400,490)
\ArrowLine(400,490)(430,520)
\DashLine(400,490)(400,460){2}
\Photon(400,460)(400,430){2}{4.5}
\ArrowLine(370,400)(400,430)
\ArrowLine(400,430)(430,400)
\Text(417,462)[]{$\Delta_{\nu}^{(2n)}$}
\Text(370,420)[]{$\psi^{(n)}$}
\Text(432,420)[]{$\bar{\psi}^{(n)}$}
\Text(370,500)[]{$\psi^{(n)}$}
\Text(432,500)[]{$\bar{\psi}^{(n)}$}
\GCirc(400,430){6}{0.9}
\Text(400,430.4)[]{$A$}
\GCirc(400,490){6}{0.9}
\Text(400,490.4)[]{$P$}

\Text(-10,310)[]{$=$}

\ArrowLine(10,370)(40,340)
\ArrowLine(40,340)(70,370)
\Photon(40,340)(40,280){2}{4.5}
\ArrowLine(10,250)(40,280)
\ArrowLine(40,280)(70,250)
\Text(55,312)[]{$U_{\mu\nu}^{(2n)}$}
\Text(10,270)[]{$\psi^{(n)}$}
\Text(72,270)[]{$\bar{\psi}^{(n)}$}
\Text(10,350)[]{$\psi^{(n)}$}
\Text(72,350)[]{$\bar{\psi}^{(n)}$}
\GCirc(40,280){6}{0.9}
\Text(40,280.4)[]{$A$}
\GCirc(40,340){6}{0.9}
\Text(40,340.4)[]{$A$}

\Text(180,200)[]{\bf Fig.\ 4}

\end{picture}

\end{center}
\newpage
\begin{center}
\begin{picture}(400,500)(0,0)
\SetWidth{0.8}

\Text(-40,460)[]{${\cal S_{C}}$}
\Text(-10,460)[]{$=$}

\ArrowLine(10,520)(40,490)
\ArrowLine(40,490)(70,520)
\Photon(40,490)(40,430){2}{4.5}
\ArrowLine(10,400)(40,430)
\ArrowLine(40,430)(70,400)
\Text(55,462)[]{$\Delta_{\mu\nu}^{(j)}$}
\Text(10,420)[]{$\psi^{(\ell)}$}
\Text(72,420)[]{$\bar{\psi}^{(\ell+j)}$}
\Text(10,500)[]{$\psi^{(i+j)}$}
\Text(72,500)[]{$\bar{\psi}^{(i)}$}
\GCirc(40,430){6}{0.9}
\Text(40,430.4)[]{$V$}
\GCirc(40,490){6}{0.9}
\Text(40,490.4)[]{$V$}

\Text(100,460)[]{$+$}

\ArrowLine(130,520)(160,490)
\ArrowLine(160,490)(190,520)
\DashLine(160,490)(160,430){2}
\ArrowLine(130,400)(160,430)
\ArrowLine(160,430)(190,400)
\Text(175,462)[]{$\Delta^{(j)}$}
\Text(130,420)[]{$\psi^{(\ell)}$}
\Text(192,420)[]{$\bar{\psi}^{(\ell+j)}$}
\Text(130,500)[]{$\psi^{(i+j)}$}
\Text(192,500)[]{$\bar{\psi}^{(i)}$}
\GCirc(160,430){6}{0.9}
\Text(160,430.4)[]{$S$}
\GCirc(160,490){6}{0.9}
\Text(160,490.4)[]{$S$}

\Text(220,460)[]{$+$}

\ArrowLine(250,520)(280,490)
\ArrowLine(280,490)(310,520)
\Photon(280,490)(280,460){2}{4.5}
\DashLine(280,460)(280,430){2}
\ArrowLine(250,400)(280,430)
\ArrowLine(280,430)(310,400)
\Text(297,462)[]{$\Delta_{\mu}^{(j)}$}
\Text(250,420)[]{$\psi^{(\ell)}$}
\Text(312,420)[]{$\bar{\psi}^{(\ell+j)}$}
\Text(250,500)[]{$\psi^{(i+j)}$}
\Text(312,500)[]{$\bar{\psi}^{(i)}$}
\GCirc(280,430){6}{0.9}
\Text(280,431)[]{$S$}
\GCirc(280,490){6}{0.9}
\Text(280,490.4)[]{$V$}

\Text(340,460)[]{$+$}

\ArrowLine(370,520)(400,490)
\ArrowLine(400,490)(430,520)
\DashLine(400,490)(400,460){2}
\Photon(400,460)(400,430){2}{4.5}
\ArrowLine(370,400)(400,430)
\ArrowLine(400,430)(430,400)
\Text(417,462)[]{$\Delta_{\nu}^{(j)}$}
\Text(370,420)[]{$\psi^{(\ell)}$}
\Text(432,420)[]{$\bar{\psi}^{(\ell+j)}$}
\Text(370,500)[]{$\psi^{(i+j)}$}
\Text(432,500)[]{$\bar{\psi}^{(i)}$}
\GCirc(400,430){6}{0.9}
\Text(400,430.4)[]{$V$}
\GCirc(400,490){6}{0.9}
\Text(400,491)[]{$S$}

\Text(-10,310)[]{$=$}

\ArrowLine(10,370)(40,340)
\ArrowLine(40,340)(70,370)
\Photon(40,340)(40,280){2}{4.5}
\ArrowLine(10,250)(40,280)
\ArrowLine(40,280)(70,250)
\Text(55,312)[]{$U_{\mu\nu}^{(j)}$}
\Text(10,270)[]{$\psi^{(\ell)}$}
\Text(72,270)[]{$\bar{\psi}^{(\ell+j)}$}
\Text(10,350)[]{$\psi^{(i+j)}$}
\Text(72,350)[]{$\bar{\psi}^{(i)}$}
\GCirc(40,280){6}{0.9}
\Text(40,280.4)[]{$V$}
\GCirc(40,340){6}{0.9}
\Text(40,340.4)[]{$V$}

\Text(180,200)[]{\bf Fig.\ 5}

\end{picture}

\end{center}

\end{document}